\documentclass[twocolumn,prl,showpacs,superscriptaddress]{revtex4-1}
\usepackage{amsmath,amsfonts,bm,natbib}
\usepackage[utf8]{inputenc}
\usepackage{graphicx}
\usepackage{bm}

\renewcommand{\vec}[1]{\bm{#1}}
\renewcommand{\epsilon}{\varepsilon}
\renewcommand{\phi}{\varphi}

\def\ket#1{\mathinner{|{#1}\rangle}}
\def\braket#1{\mathinner{\langle{#1}\rangle}}

\begin{document}
\title{Microwave--Induced Fano--Feshbach Resonances}
\author{D.J. Papoular}
\email{david.papoular@lptms.u-psud.fr} 
\affiliation{\mbox{
    Laboratoire de Physique Th\'eorique et Mod\`eles Statistiques, 
    CNRS, Universit\'e Paris-Sud, 91405, Orsay,
    France}} 
\author{G.V. Shlyapnikov} 
\affiliation{\mbox{
    Laboratoire de Physique Th\'eorique et Mod\`eles Statistiques, 
    CNRS, Universit\'e Paris-Sud, 91405, Orsay, 
    France}}
\affiliation{\mbox{
    Van der Waals-Zeeman Institute, 
    University of Amsterdam, Valckenierstraat 65/67,
    1018 XE Amsterdam, The Netherlands}}
\author{J. Dalibard}
\affiliation{Laboratoire Kastler Brossel, CNRS, UPMC, 
             Ecole Normale Sup\'erieure,
             24 rue Lhomond, 75231, Paris, France}
\date{\today}

\begin{abstract}
We investigate the possibility to control
the s--wave scattering length for the interaction
between cold bosonic atoms by using a microwave field.
Our scheme applies to any atomic species with a ground state that 
is split by hyperfine interaction. We discuss more specifically the case 
of alkali atoms and calculate the change in the scattering  length for
${}^7\mathrm{Li}$, ${}^{23}\mathrm{Na}$, ${}^{41}\mathrm{K}$,
${}^{87}\mathrm{Rb}$, and ${}^{133}\mathrm{Cs}$. 
Our results yield optimistic prospects for experiments with the four
latter species.
\end{abstract}
\maketitle
\newcommand{\Eb}{E_{\rm b}} \newcommand{\Eth}{E_{\rm th}}
Cold atomic gases constitute model systems to investigate a wealth of
collective quantum phenomena, ranging from few--body physics
\cite{kohler:2006,braaten:2006} to condensed matter
problems \cite{bloch:2008,giorgini:2008}. In particular one can control 
the strength of the interparticle interactions, using scattering
resonances that occur in 
a collision between two atoms with low energy. These so--called
Fano--Feshbach resonances (FFRs) arise when the entrance collision
channel, with an energy threshold $\Eth$, is coupled to another
channel that supports a molecular bound state $b$ at an energy $\Eb$
close to $\Eth$ 
\cite{kohler:2006,chin:2008,moerdijk:1995,tiesinga:1993}.
The scattering length
that characterizes the s-wave scattering between the two atoms has a
dispersive variation with $\Eth-\Eb$, and can in principle be tuned
to a value with arbitrary sign and magnitude.

In practice FFRs are generally obtained by adjusting the external
magnetic field. One takes advantage of the degenerate structure of the
lowest electronic energy level of the atoms. In the case of alkali
atoms that are widely used in cold atom experiments, the degeneracy
emerges from the spins of the valence electron and of the nucleus. If
the magnetic moment of the bound level $b$ is different from that of
the entrance channel, the energy difference $\Eth-\Eb$ can be tuned by
scanning the external field. This leads to a resonant variation of the
scattering length, with a width that depends on the coupling between
the two channels and, hence, on the details of the interaction 
between the colliding atoms. For some atomic species, such as lithium,
potassium, or cesium, these magnetic FFRs have been an invaluable tool
for many studies related to atom--atom interactions 
\cite{kohler:2006,braaten:2006,bloch:2008,giorgini:2008}. However, the
absence of external control on the width of magnetic FFRs and their
occurence only for fixed values of the magnetic field may
constitute a serious drawback. For sodium atoms ($^{23}$Na) for
example, the identified resonances are in the 1000~G region with a
width around 1~G or less \cite{inouye:1998,stenger:1999}.  Similar
values are found for polarized rubidium atoms ($^{87}$Rb)
\cite{marte:2002}. These large field values and narrow widths severely
limit the use of FFRs for these species.

In this Rapid Communication we study an alternative to magnetic FFRs, where the
entrance channel is resonantly coupled by a microwave (mw) field to a
bound state in another collision channel. All relevant states
correspond to the electronic ground level of the atoms, and
the resonance is reached by adjusting the frequency of the mw.
The width of the resonance is related to the strength of the magnetic
dipole coupling between the two channels and is proportional to the mw
intensity. Our scheme is reminiscent of optical FFRs, as
proposed in \cite{fedichev:1996} and
experimentally demonstrated in \cite{theis:2004,enomoto:2008}. 
There, the bound state $b$ was an electronically
excited dimer. Although optical FFRs, which rely on electric rather
than magnetic dipole coupling, allow in principle stronger resonances,
their practical use is limited by the unavoidable losses due to
spontaneous emission processes. One can also use a pair of laser beams 
to coherently couple two states from the ground electronic level
\cite{bohn:1999}. However,
for a given change of the scattering length this method leads to a similar 
spontaneous emission rate as in the case of a single-photon excitation
\cite{thalhammer:2005}.  
So far the lifetimes of atomic samples
submitted to optical FFR were limited to tens of milliseconds,
which is likely to
be too short to reach a many--body equilibrium state. By contrast
mw--FFRs do not suffer from any spontaneous emission process
and the associated loss rates should be comparable to those observed
with magnetic FFRs.

So far the use of mw or radiofrequency (rf) fields has been discussed 
in relation to manipulating existing FFRs
\cite{thompson:2005,zhang:2009,kaufman:2009}. 
Zhang {\it et al} \cite{zhang:2009}
proposed to consider magnetic FFRs for atomic states dressed by a
two--color Raman process or by a rf field. The idea was to provide
an independent control of different scattering lengths in
multicomponent gases. The rf coupling of several magnetic FFRs
has been studied experimentally and theoretically in
\cite{kaufman:2009}. The analysis showed that the main role of rf
is to couple the bound states that give rise to these resonances. Our
idea of inducing new FFRs by using mw fields brings in a novel
physical context. We focus on the case of zero static magnetic field,
which is presently put forward in the studies of ground--state
properties and quantum phase transitions in spinor Bose gases.
Such experiments require extremely low magnetic fields
($<$10 mG), and the manipulation of the interatomic interactions 
becomes crucial for the observation of quantum transitions
and their dynamics. Our scheme is also different from \cite{thompson:2005} 
where a resonant oscillating magnetic field was used to enhance the 
production of diatomic molecules
near an existing FFR. In our case the bound state that is coupled to the 
entrance channel is only virtually populated, and no molecule is produced 
in the collision.  

\begin{figure}
  \begin{center}
    \includegraphics[width=.7\linewidth]{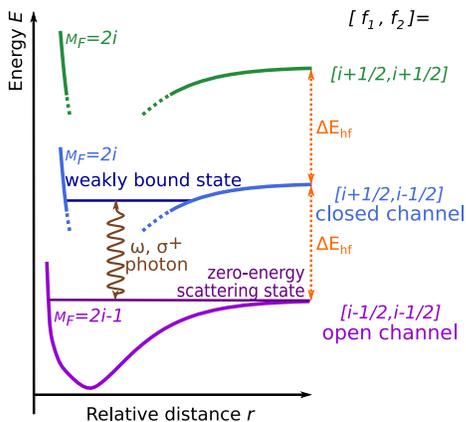}
    \caption{\label{fig:twotothree} 
Fano-Feshbach resonance in a collision between two atoms, 
induced by an oscillatory magnetic field. If the mw frequency $\omega$ 
approaches the energy difference between the
incident scattering state  and a weakly--bound dimer state, 
the pair of atoms undergoes virtual
spin--flip transitions which cause a resonant variation of the
scattering length with $\omega$.}
  \end{center}
\end{figure}
\indent
For simplicity we study in the following a collision between two
identical bosonic atoms prepared in the same internal state. Our
treatment can be straightforwardly extended to fermionic particles,
and to mixtures of atoms in different internal states. More
specifically, we consider alkali atoms whose ground level is split by
the hyperfine interaction into two sublevels with total spins
$f_+=i+1/2$ and $f_-=i-1/2$, where $i$ is the nuclear spin. The
frequency $\omega$ of the mw field is chosen close to
the hyperfine splitting $\Delta E_\mathrm{hf}$ between these two
sublevels
(see Fig.~\ref{fig:twotothree}). The various
collision channels can then be grouped into 3 categories
corresponding to asymptotic states with (i) both atoms in $f_+$,
(ii) one atom in $f_+$ and one in $f_-$, (iii) both atoms in $f_-$.
We consider in the following the case of a $f_-f_-$ collision, and the
mw induces a quasi--resonant transition to a bound state in a potential
from the $f_+ f_-$ group as shown in Fig.~\ref{fig:twotothree}.

We describe the system in the center--of--mass frame of the atom pair.
Neglecting the weak coupling between the atomic spins, the atom--atom
interaction is spatially isotropic. We limit our analysis to s--wave
collisions governed by the radial Hamiltonian (see
\cite{verhaar:2009} and refs. in)
\begin{equation} \label{eq:totham}
  H=\frac{p^2}{2\mu}+V_\mathrm{c}(r)+V_\mathrm{hf} 
    +\hbar \omega\, a^\dagger a+ W
   = H_0 + W,
\end{equation}
where $r$ is the interatomic distance, $p$ is its conjugate
momentum, and $\mu=m/2$ is the reduced mass of the atom pair.  The
central part $V_\mathrm{c}(r)$ of the interaction is given by
$V_\mathrm{c}(r)=V_S(r)P_S+V_T(r)P_T$, where $P_S$ and $P_T$ are the
projection operators onto the electronic--singlet and triplet subspaces.
The term
$V_\mathrm{hf}=a_\mathrm{hf}(\vec{s}_1\cdot\vec{i}_1+\vec{s}_2\cdot\vec{i}_2)$
is the hyperfine interaction, where $\vec{s}_j$ and $\vec{i}_j$
stand for the spin operators of the electron and 
nucleus of atom $j$.  We use a quantum description for the mw field and
$a^\dagger$ is the creation operator for a mw photon in the relevant
mode. The magnetic dipole interaction between the atoms and the mw is
$W=-\vec{M}\cdot\vec{B}$, where $\vec{M}$ is the total magnetic dipole
operator of the atom pair, and
$\vec{B}={b_0}(\vec{\epsilon}
a+\vec{\epsilon}^* a^\dagger)/\sqrt{2}$
is the magnetic field operator for the
mode of polarization $\vec{\epsilon}$.  As usual in the dressed--atom
approach \cite{cohen:1992}, the amplitude $b_0$ and the number of
photons $N$ in the mw mode are arbitrary. The only relevant
physical quantity is the amplitude of the applied mw field
$B_0=b_0\sqrt{N}$ (with $N \gg 1$).
We assume that the
magnetic field is $\sigma^+$--polarized with respect to the quantization
axis $\vec{e}_z$ \footnote{This restriction leads to simpler algebra, 
but is not essential: any polarization can be decomposed 
into $\sigma_\pm$ components and, for a given $\omega$, 
only one of the components will induce the desired resonant coupling 
to a bound state.}.
The valence electron in each atom has zero angular momentum and 
$W$ reduces
to \footnote{
  In Eq.~\ref{eq:couplingW}
  we omit a small coupling of the mw to the nuclear spins
  which does not affect the results.
  }:
\begin{equation}
  \label{eq:couplingW}
  W=W_1\,(S^+\, a + S^-\, a^\dagger),
\end{equation}
where $W_1\!=\!\mu_\mathrm{B} b_0/\hbar$, $\mu_\mathrm{B}$ is the Bohr magneton,
and $S^{\pm}\!=\!S_x\!\pm\! iS_y$, with
$\vec{S}\!=\!\vec{s}_1\!+\!\vec{s}_2$
being the total electron spin.

We study the scattering properties of $H$
using two different methods: 
(i) if the mw Rabi frequency $\mu_{\rm B}B_0$ is much smaller 
than the binding energy $|E_T|$ of the dimer and the 
level spacing in the closed channel, 
the scattering is well
described by a single--resonance two--channel model;
(ii) for $\mu_{\rm B}B_0\gtrsim |E_T|$, a more general description
is obtained through a full coupled--channel calculation.

We first describe method (i).
We consider $H$ as a two--channel model
\cite{kohler:2006}
where $H_0$ is the bare Hamiltonian and $W$ is the coupling operator.
The symmetries of
$H_0$ allow the choice of bare open-- and closed--channel
wavefunctions which have well--defined photon numbers $N$,
total spin $F$,
and total spin projection $M_F$ along the quantization axis
($\vec{F}=\vec{s}_1+\vec{i}_1+\vec{s}_2+\vec{i}_2$), whereas $W$
directly couples subspaces with $\Delta M_F=-\Delta N =\pm 1$.
The coupling term
$W$ does not vanish in the limit of infinitely separated atoms.
Hence, there is a difference 
$\Delta$ in the scattering threshold energy of $H$ compared to
that of $H_0$. If the detuning $\delta$ of the mw with respect
to the single atom hyperfine splitting $\Delta E_\mathrm{hf}$
is greater than $\mu_\mathrm{B} B_0$, 
then $\Delta\sim(\mu_\mathrm{B} B_0)^2/\delta$.
Method (i) is applicable when $\Delta\ll |E_T|$ and can be neglected.
Near resonance, where $|\delta|\sim |E_T|$, this
condition requires $\mu_\mathrm{B} B_0\ll |E_T|$. 

We start by stating a selection rule associated with $W$.
All internal states in the $M_F=2i+1$ and $M_F=2i$ subspaces are
electronic--triplet states. More precisely, the $M_F=2i+1$ subspace
has dimension one, with $\ket{S=1,I=2i,F=2i+1,M_F=2i+1}$ as a basis
vector ($I$ determines the modulus of the total nuclear spin
$\vec{I}=\vec{i}_1+\vec{i}_2$).
The $M_F=2i$ subspace has dimension two
\footnote{For $s$--wave collisions between bosons, only symmetric
internal states are relevant.},
and it is spanned by
$\ket{\eta_1}=\ket{S=1,I=2i,F=2i,M_F=2i}$ and
$\ket{\eta_2}=\ket{S=1,I=2i,F=2i+1,M_F=2i}$.
The spatial components of the eigenfunctions of $H_0$ in these
subspaces decouple from the internal states and are all
eigenfunctions of the triplet Hamiltonian $H_T=p^2/2\mu+V_T(r)$. The
bare open-- and closed--channel spatial wavefunctions are thus
orthogonal. The operator $W$ does not act on the spatial parts of the
wavefunctions. Hence, its matrix element between an open--channel state with
$M_F=2i$ and a bound state with $M_F=2i+1$ is zero.
Therefore, $W$ cannot induce any resonance between these two subspaces.

We now consider a resonance between the $M_F=2i-1$ subspace 
(dimension five) and the $M_F=2i$
subspace. 
For the bare open--channel wavefunction we choose the $M_F=2i-1$
threshold--energy scattering state
$\ket{\Psi^{(2i-1)}_{\vec{k}=\vec{0}}}$, in the
presence of $N$ photons.  
For large interatomic separations,
this state corresponds to the two--particle state
$\ket{f_-f_-,F=2i-1,M_F=2i-1}$ in which both atoms have 
$f=m_f=i-1/2$ (see Fig.~\ref{fig:twotothree}).
The bare closed--channel wavefunction is chosen in the 
form $\ket{\Psi^{(2i)}_0}=\ket{\phi_T,\eta_1}$, where $\phi_T(r)$ is a
bound state of $H_T$, and $\ket{\eta_1}$ is defined above 
(see Fig.~\ref{fig:twotothree}). 

The single-resonance two--channel model leads to the usual behavior for the
scattering length as a function of the frequency $\omega$
close to a FFR resonance:
\begin{equation}
  \label{eq:ares}
  a(\omega)
  =a_\mathrm{bg}
  \left(1+ \frac{\Delta \omega}
    {\omega-\omega_\mathrm{res}}
  \right) \ .
\end{equation} 
The background scattering length
$a_\mathrm{bg}$ corresponds to a collision in the absence of mw, between
two atoms in the state $|f=i-1/2,m_f=i-1/2\rangle$. 
The resonance position is given by
$\hbar \omega_\mathrm{res}\approx\Delta E_\mathrm{hf}-|E_T|+\alpha B_0^2$,
where $\alpha B_0^2$
is a small shift due to the coupling between the open and closed
channels \cite{kohler:2006}.
The width $\Delta \omega$ of the mw FFR is: 
\begin{equation}
  \hbar\Delta \omega=
  \frac{1}{2\pi}\,
  \frac{\mu}{a_\mathrm{bg}\hbar^2}\,
  (\mu_{{\rm B}}B_0)^2\,
  |\braket{\Psi_0^{(2i)}|S^+|\Psi_{\vec{k}=\vec{0}}^{(2i-1)}}|^2\ . 
  \label{eq:width}
\end{equation}
It is proportional to 
the mw intensity $B_0^2$ and to the spin--flip Franck--Condon factor
$|\braket{\Psi_0^{(2i)}|S^+|\Psi_{\vec{k}=\vec{0}}^{(2i-1)}}|^2$.

For a given atomic species, method (i) requires the calculation
of
$\Psi^{(2i-1)}_{\vec{k}=\vec{0}}(r)$ and $\phi_T(r)$.
We
account for the spin--recoupling phenomenon
\cite{kohler:2006} through the coupled--channel method
\cite{stoof:1988}, encode the short--range physics in the
accumulated--phase boundary condition
\cite{verhaar:1993,verhaar:2009}, and use the relaxation method  
\cite{nr3:2007}
to solve the resulting two--point boundary--value differential 
systems \footnote{The same approaches were used for method (ii).}.

We have performed calculations for ${}^7\mathrm{Li}$,
${}^{23}\mathrm{Na}$, ${}^{41}\mathrm{K}$, ${}^{87}\mathrm{Rb}$,
and ${}^{133}\mathrm{Cs}$.
We use the hyperfine splittings reported in \cite{arimondo:1977}
and the singlet and triplet potentials from
\cite{barakat:1986,yan:1996,colavecchia:2003,
vanabeelen:1999,
falke:2008,
marte:2002,kokkelmans:2008,
leo:2000,amiot:2002,chin:2004,vanhaecke:2004,xie:2009}.
The accumulated--phase boundary condition is applied at the radii
 $r_0=10\,\mathrm{a_0}$ for ${}^{7}\mathrm{Li}$,
$r_0=16\,\mathrm{a}_0$ for ${}^{23}\mathrm{Na}$, ${}^{41}\mathrm{K}$
and ${}^{87}\mathrm{Rb}$, and
$r_0=20\,\mathrm{a_0}$ for ${}^{133}\mathrm{Cs}$.
We calculate the initial
phases of the zero--energy scattering wavefunctions at $r_0$ through
back--integration using the singlet and triplet scattering lengths
\cite{abraham:1997,colavecchia:2003,
vanabeelen:1999,
falke:2008,
marte:2002,
chin:2004}.
The energy derivatives of these phases are taken from
\cite{verhaar:2009,vanabeelen:1999} for ${}^{87}\mathrm{Rb}$ and
${}^{23}\mathrm{Na}$, and are calculated for the other species
using the triplet and singlet potentials.
Our results are
given in Table~\ref{tab:poswidths}. In practice we find that the broadest resonance widths $\Delta\omega$,
as given  by Eq. \ref{eq:width}, are obtained  by choosing $\phi_T(r)$
as  the  highest  bound  state  of the  triplet  potential.   For  all
considered atomic  species except ${}^{133}\mathrm{Cs}$, this is  the
resonance we report in Table \ref{tab:poswidths}.  However,  in the
case  of ${}^{133}\mathrm{Cs}$,
the  highest--energy bound  state  is so  weakly bound  ($|E_T|=h\cdot
5\,\mathrm{kHz}$)    that    the     hyperbolic    behavior    of  $a$
(Eq. \ref{eq:ares}) is not valid for $B_0\gtrsim 1\,\mathrm{mG}$,
and we therefore
report the resonance  obtained with the second--highest bound  state of $V_T$
($|E_T|=h\cdot 110\,\mathrm{MHz}$). 
\begin{table}
  \begin{center}
    \begin{tabular}{|l|c|c|c|c|c|}
      \hline
      & \bm{${}^7\mathrm{Li}$}  
      & \bm{${}^{23}\mathrm{Na}$}
      & \bm{${}^{41}\mathrm{K}$}
      & \bm{${}^{87}\mathrm{Rb}$}
      & \bm{${}^{133}\mathrm{Cs}$}\\
      \hline
      \bm{$|E_T|/h$} ($\mathrm{MHz}$)
      & 12000 & 200 & 140 & 25 & 110\\
      \hline
      \bm{$\omega_\mathrm{res}/2\pi$} ($\mathrm{GHz}$)
      & $12$   & $1.6$ & $0.12$ & $6.8$ & $9.1$\\
      \hline
      \bm{$\alpha$} ($\mathrm{kHz/G^2}$)
      & $0.33$ & $6.8$& $21$ & $120$ & $30$\\
      \hline
      \bm{$\Delta\omega/2\pi$} ($\mathrm{Hz}$)
      &  $6$  & $1400$  & $350$ & $60$  & $-4500$ \\
      \hline
    \end{tabular}
  \end{center}
\caption{\label{tab:poswidths}
  Characteristics of the mw--FFR in ${}^7\mathrm{Li}$, ${}^{23}\mathrm{Na}$,
  ${}^{41}\mathrm{K}$, ${}^{87}\mathrm{Rb}$, and ${}^{133}\mathrm{Cs}$ involving
  the triplet bound states with energies $E_T$.
  The width $\Delta\omega$ scales as $B_0^2$ and is given for
  $B_0=1\,\mathrm{G}$.}
\end{table}

The largest resonance width is obtained for ${}^{133}\,\mathrm{Cs}$
($-4500\,\mathrm{Hz}$ for $B_0=1\,\mathrm{G}$).
Relatively large widths are also obtained for $^{23}$Na and $^{41}$K.
In the case of
${}^{87}\mathrm{Rb}$, the singlet and triplet scattering lengths
differ by less than $10\,\%$ \cite{marte:2002}. Hence, the
near--threshold properties of the singlet and triplet Hamiltonians are
similar.  The open-- and closed--channel wavefunctions
are thus nearly orthogonal, which leads to a reduction of their overlap 
and to a narrower resonance.
For ${}^7\mathrm{Li}$ the triplet scattering length is negative
\cite{abraham:1997}, and $H_T$ does not support weakly bound molecular states
\cite{kohler:2006}. The last bound state has a small spatial extent,
which leads to an even narrower resonance.

We now turn to method (ii), where  we take into account that 
colliding atoms are asymptotically in  dressed atomic states. 
This method can be used for larger Rabi frequencies, such that
$\mu_{\rm B}B_0\gtrsim \Delta$.  We restrict
the  full Hamiltonian  $H$ to  
the eight--dimensional subspace
spanned  by all  internal
states  in  the  $M_F=2i+1$, $2i$,  and  $2i-1$  subspaces. 
For  given values of $B_0$ and $\omega$, we calculate
the eight--component scattering state $\ket{\Psi_{B_0,\omega}}$ of $H$ 
corresponding, for large interatomic separations, to dressed-state
atoms with zero kinetic  energy. The  scattering length $a(B_0,\omega)$  is 
extracted from  the asymptotic behavior  of this wavefunction.   For a
given $\mu_\mathrm{B}B_0\ll |E_T|$, we have checked that we recover the
hyperbolic behavior of Eq. \ref{eq:ares}.   We  have also evaluated 
the coefficient   $\alpha$  giving  the   shift  of   the  resonance   
position $\omega_\mathrm{res}$ (see Table \ref{tab:poswidths}).

\begin{figure*}
  \begin{center}
    \includegraphics[width=.20\linewidth,angle=-90]{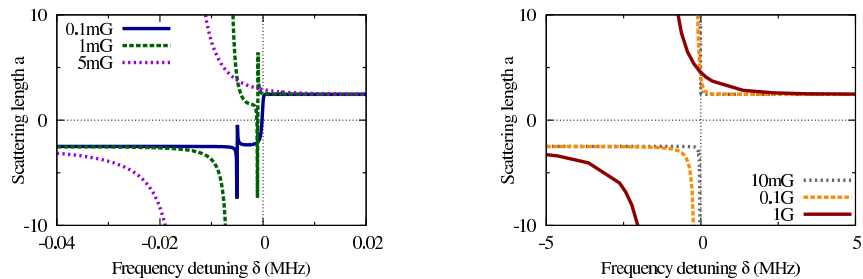}
    \caption{\label{fig:res_cesium_highest} 
      Resonance in ${}^{133}\mathrm{Cs}$ involving the
      bound state with energy
      $|E_T|=h\cdot 5\,\mathrm{kHz}
            =\mu_\mathrm{B}\cdot 4\mathrm{mG}$,
      for $B_0$ ranging from $0.1\,\mathrm{mG}$ to $1\,\mathrm{G}$,
      calculated using method (ii). The scattering length $a$ is
      expressed in units of $1000\,\mathrm{a}_0$.} 
    \end{center}
\end{figure*}
Method (ii) allows us to investigate the mw--FFR in
${}^{133}\mathrm{Cs}$
involving the highest bound state of 
$V_T$, where Eq.~\ref{eq:ares} is not applicable for 
$B_0\gtrsim 1\,\mathrm{mG}$. In order to avoid
inelastic processes we assume
that the atoms are asymptotically in the lowest atomic dressed state.
For large $\delta<0$, this state
corresponds to the two--particle state 
$\ket{f_-f_-,F=2i-1,M_F=2i-1}$,
and therefore $a=a_\mathrm{bg}=-2500\,\mathrm{a_0}$.
For large $\delta>0$ it corresponds to $\ket{f_+f_+,F=2i+1,M_F=2i+1}$,
with $a=a_T=2400\,\mathrm{a_0}$.
For $B_0\lesssim 1$~mG the resonance is hyperbolic, as predicted
by method (i) (see Eq.~\ref{eq:ares}). 
For larger $B_0$
the scattering length becomes very large for  
$\hbar\omega\approx\Delta E_\mathrm{hf}$, 
but $a(\omega)$ no longer 
satisfies Eq.~\ref{eq:ares}.
Figure~\ref{fig:res_cesium_highest}
shows how the dependence $a(\omega)$ evolves when $B_0$ increases from
$0.1\,\mathrm{mG}$ to $1\,\mathrm{G}$. 
In the $\delta>0$ region, collisions between atoms in the
``stretched'' state $f_+=m_f=i+1/2=4$ occur with a large inelastic
rate because of dipole--dipole interactions \cite{soding:1998}.
Therefore, one should operate in the $\delta<0$ region,
where the contamination of the collision state by the stretched state
is small.
A detailed modeling of the
large--$B_0$ FFRs
will be presented elsewhere \cite{papoular:tbp}.
 
Our results draw optimistic prospects for modifying the scattering length
in atomic gases using a microwave field. Using small resonant
transmitting loop  antennas in the near--field regime,
it is possible to reach mw magnetic field 
amplitudes $B_0\sim 10$~G in the desired frequency range, while
keeping a reasonable incident electromagnetic power (below 10 W). 
The resonance widths obtained for the hyperbolic resonances in all
atomic species except ${}^7\mathrm{Li}$ are then well above 1~mG, 
and thus notably exceed typical magnetic field fluctuations 
in setups with an efficient magnetic shielding. The
non--hyperbolic resonance obtained with ${}^{133}\mathrm{Cs}$ has
a width of the order of $1\,\mathrm{G}$ for $B_0=1\,\mathrm{G}$.
Our scheme can be readily transposed to fermionic atoms,
multicomponent gases, and heteronuclear mixtures,
and it can allow for a fine tuning of interspecies 
interactions in all three cases.  

We thank M. K\"ohl, S. Kokkelmans, F. Gerbier and Z. Hadzibabic 
for helpful discussions. This work is supported by R{\'e}gion Ile de 
France IFRAF, by ANR (Grant ANR-08-BLAN-65 BOFL), by the EU project SCALA,
and by the Dutch Foundation FOM. LPTMS is a mixed research unit No.8626 of 
CNRS and Universit\'e Paris Sud. LKB is a mixed research unit No.8552
of CNRS, ENS, and Universit{\'e} Pierre et Marie Curie. 

%\bibliography{mwffr}

\begin{thebibliography}{10}%
\makeatletter
\providecommand \@ifxundefined [1]{%
 \ifx #1\undefined \expandafter \@firstoftwo
 \else \expandafter \@secondoftwo
\fi
}%
\providecommand \@ifnum [1]{%
 \ifnum #1\expandafter \@firstoftwo
 \else \expandafter \@secondoftwo
\fi
}%
\providecommand \enquote [1]{``#1''}%
\providecommand \bibnamefont  [1]{#1}%
\providecommand \bibfnamefont [1]{#1}%
\providecommand \citenamefont [1]{#1}%
\providecommand\href[0]{\@sanitize\@href}%
\providecommand\@href[1]{\endgroup\@@startlink{#1}\endgroup\@@href}%
\providecommand\@@href[1]{#1\@@endlink}%
\providecommand \@sanitize [0]{\begingroup\catcode`\&12\catcode`\#12\relax}%
\@ifxundefined \pdfoutput {\@firstoftwo}{%
 \@ifnum{\z@=\pdfoutput}{\@firstoftwo}{\@secondoftwo}%
}{%
 \providecommand\@@startlink[1]{\leavevmode}%
 \providecommand\@@endlink[0]{}%
}{%
 \providecommand\@@startlink[1]{%
  \leavevmode
  \pdfstartlink
   attr{/Border[0 0 1 ]/H/I/C[0 1 1]}%
   user{/Subtype/Link/A<</Type/Action/S/URI/URI(#1)>>}%
  \relax
 }%
 \providecommand\@@endlink[0]{\pdfendlink}%
}%
\providecommand \url  [0]{\begingroup\@sanitize \@url }%
\providecommand \@url [1]{\endgroup\@href {#1}{\urlprefix}}%
\providecommand \urlprefix [0]{URL }%
\providecommand \Eprint[0]{\href }%
\@ifxundefined \urlstyle {%
  \providecommand \doi [1]{doi:\discretionary{}{}{}#1}%
}{%
  \providecommand \doi [0]{doi:\discretionary{}{}{}\begingroup
  \urlstyle{rm}\Url }%
}%
\providecommand \doibase [0]{http://dx.doi.org/}%
\providecommand \Doi[1]{\href{\doibase#1}}%
\providecommand \bibAnnote [3]{%
  \BibitemShut{#1}%
  \begin{quotation}\noindent
    \textsc{Key:}\ #2\\\textsc{Annotation:}\ #3%
  \end{quotation}%
}%
\providecommand \bibAnnoteFile [2]{%
  \IfFileExists{#2}{\bibAnnote {#1} {#2} {\input{#2}}}{}%
}%
\providecommand \typeout [0]{\immediate \write \m@ne }%
\providecommand \selectlanguage [0]{\@gobble}%
\providecommand \bibinfo [0]{\@secondoftwo}%
\providecommand \bibfield [0]{\@secondoftwo}%
\providecommand \translation [1]{[#1]}%
\providecommand \BibitemOpen[0]{}%
\providecommand \bibitemStop [0]{}%
\providecommand \bibitemNoStop [0]{.\EOS\space}%
\providecommand \EOS [0]{\spacefactor3000\relax}%
\providecommand \BibitemShut [1]{\csname bibitem#1\endcsname}%
%</preamble>
\bibitem{kohler:2006}%
  \BibitemOpen
  \bibfield{author}{%
  \bibinfo {author} {\bibfnamefont{T.}~\bibnamefont{K{\"o}hler}}
  \emph{et~al.},\ }%
  \bibfield{journal}{%
  \bibinfo {journal} {Rev. Mod. Phys.}\ }%
  \textbf{\bibinfo {volume} {78}},\ \bibinfo {pages} {1311} (\bibinfo {year}
  {2006})%
  \bibAnnoteFile{NoStop}{kohler:2006}%
\bibitem{braaten:2006}%
  \BibitemOpen
  \bibfield{author}{%
  \bibinfo {author} {\bibfnamefont{E.}~\bibnamefont{Braaten}}\ and\ \bibinfo
  {author} {\bibfnamefont{H.-W.}\ \bibnamefont{Hammer}},\ }%
  \bibfield{journal}{%
  \bibinfo {journal} {Phys. Rep.}\ }%
  \textbf{\bibinfo {volume} {428}},\ \bibinfo {pages} {259} (\bibinfo {year}
  {2006})%
  \bibAnnoteFile{NoStop}{braaten:2006}%
\bibitem{bloch:2008}%
  \BibitemOpen
  \bibfield{author}{%
  \bibinfo {author} {\bibfnamefont{I.}~\bibnamefont{Bloch}} \emph{et~al.},\ }%
  \bibfield{journal}{%
  \Doi{10.1103/RevModPhys.80.885}{\bibinfo {journal} {Rev. Mod. Phys.}}\ }%
  \textbf{\bibinfo {volume} {80}},\ \bibinfo {eid} {885} (\bibinfo {year}
  {2008})%
  \bibAnnoteFile{NoStop}{bloch:2008}%
\bibitem{giorgini:2008}%
  \BibitemOpen
  \bibfield{author}{%
  \bibinfo {author} {\bibfnamefont{S.}~\bibnamefont{Giorgini}} \emph{et~al.},\
  }%
  \bibfield{journal}{%
  \Doi{10.1103/RevModPhys.80.1215}{\bibinfo {journal} {Rev. Mod. Phys.}}\ }%
  \textbf{\bibinfo {volume} {80}},\ \bibinfo {eid} {1215} (\bibinfo {year}
  {2008})%
  \bibAnnoteFile{NoStop}{giorgini:2008}%
\bibitem{chin:2008}%
  \BibitemOpen
  \bibfield{author}{%
  \bibinfo {author} {\bibfnamefont{C.}~\bibnamefont{Chin}} \emph{et~al.}}%
   (\bibinfo {year} {2008}),\
  \Eprint{http://arxiv.org/abs/0812.1496}{arXiv:0812.1496}%
  \bibAnnoteFile{NoStop}{chin:2008}%
\bibitem{moerdijk:1995}%
  \BibitemOpen
  \bibfield{author}{%
  \bibinfo {author} {\bibfnamefont{A.~J.}\ \bibnamefont{Moerdijk}}
  \emph{et~al.},\ }%
  \bibfield{journal}{%
  \Doi{10.1103/PhysRevA.51.4852}{\bibinfo {journal} {Phys. Rev. A}}\ }%
  \textbf{\bibinfo {volume} {51}},\ \bibinfo {pages} {4852} (\bibinfo {year}
  {1995})%
  \bibAnnoteFile{NoStop}{moerdijk:1995}%
\bibitem{tiesinga:1993}%
  \BibitemOpen
  \bibfield{author}{%
  \bibinfo {author} {\bibfnamefont{E.}~\bibnamefont{Tiesinga}} \emph{et~al.},\
  }%
  \bibfield{journal}{%
  \Doi{10.1103/PhysRevA.47.4114}{\bibinfo {journal} {Phys. Rev. A}}\ }%
  \textbf{\bibinfo {volume} {47}},\ \bibinfo {pages} {4114} (\bibinfo {year}
  {1993})%
  \bibAnnoteFile{NoStop}{tiesinga:1993}%
\bibitem{inouye:1998}%
  \BibitemOpen
  \bibfield{author}{%
  \bibinfo {author} {\bibfnamefont{S.}~\bibnamefont{Inouye}} \emph{et~al.},\ }%
  \bibfield{journal}{%
  \bibinfo {journal} {Nature}\ }%
  \textbf{\bibinfo {volume} {392}},\ \bibinfo {pages} {151} (\bibinfo {year}
  {1998})%
  \bibAnnoteFile{NoStop}{inouye:1998}%
\bibitem{stenger:1999}%
  \BibitemOpen
  \bibfield{author}{%
  \bibinfo {author} {\bibfnamefont{J.}~\bibnamefont{Stenger}} \emph{et~al.},\
  }%
  \bibfield{journal}{%
  \bibinfo {journal} {Phys. Rev. Lett.}\ }%
  \textbf{\bibinfo {volume} {82}},\ \bibinfo {pages} {2422} (\bibinfo {year}
  {1999})%
  \bibAnnoteFile{NoStop}{stenger:1999}%
\bibitem{marte:2002}%
  \BibitemOpen
  \bibfield{author}{%
  \bibinfo {author} {\bibfnamefont{A.}~\bibnamefont{Marte}} \emph{et~al.},\ }%
  \bibfield{journal}{%
  \Doi{10.1103/PhysRevLett.89.283202}{\bibinfo {journal} {Phys. Rev. Lett.}}\
  }%
  \textbf{\bibinfo {volume} {89}},\ \bibinfo {pages} {283202} (\bibinfo {year}
  {2002})%
  \bibAnnoteFile{NoStop}{marte:2002}%
\bibitem{fedichev:1996}%
  \BibitemOpen
  \bibfield{author}{%
  \bibinfo {author} {\bibfnamefont{P.~O.}\ \bibnamefont{Fedichev}}
  \emph{et~al.},\ }%
  \bibfield{journal}{%
  \bibinfo {journal} {Phys. Rev. Lett.}\ }%
  \textbf{\bibinfo {volume} {77}},\ \bibinfo {pages} {2913} (\bibinfo {year}
  {1996})%
  \bibAnnoteFile{NoStop}{fedichev:1996}%
\bibitem{theis:2004}%
  \BibitemOpen
  \bibfield{author}{%
  \bibinfo {author} {\bibfnamefont{M.}~\bibnamefont{Theis}} \emph{et~al.},\ }%
  \bibfield{journal}{%
  \Doi{10.1103/PhysRevLett.93.123001}{\bibinfo {journal} {Phys. Rev. Lett.}}\
  }%
  \textbf{\bibinfo {volume} {93}},\ \bibinfo {pages} {123001} (\bibinfo {year}
  {2004})%
  \bibAnnoteFile{NoStop}{theis:2004}%
\bibitem{enomoto:2008}%
  \BibitemOpen
  \bibfield{author}{%
  \bibinfo {author} {\bibfnamefont{K.}~\bibnamefont{Enomoto}} \emph{et~al.},\
  }%
  \bibfield{journal}{%
  \bibinfo {journal} {Phys. Rev. Lett.}\ }%
  \textbf{\bibinfo {volume} {101}},\ \bibinfo {pages} {203201} (\bibinfo {year}
  {2008})%
  \bibAnnoteFile{NoStop}{enomoto:2008}%
\bibitem{bohn:1999}%
  \BibitemOpen
  \bibfield{author}{%
  \bibinfo {author} {\bibfnamefont{J.}~\bibnamefont{Bohn}}\ and\ \bibinfo
  {author} {\bibfnamefont{P.}~\bibnamefont{Julienne}},\ }%
  \bibfield{journal}{%
  \bibinfo {journal} {Phys. Rev. A}\ }%
  \textbf{\bibinfo {volume} {60}},\ \bibinfo {pages} {414} (\bibinfo {year}
  {1999})%
  \bibAnnoteFile{NoStop}{bohn:1999}%
\bibitem{thalhammer:2005}%
  \BibitemOpen
  \bibfield{author}{%
  \bibinfo {author} {\bibfnamefont{G.}~\bibnamefont{Thalhammer}}
  \emph{et~al.},\ }%
  \bibfield{journal}{%
  \bibinfo {journal} {Phys. Rev. A}\ }%
  \textbf{\bibinfo {volume} {71}},\ \bibinfo {pages} {033403} (\bibinfo {year}
  {2005})%
  \bibAnnoteFile{NoStop}{thalhammer:2005}%
\bibitem{thompson:2005}%
  \BibitemOpen
  \bibfield{author}{%
  \bibinfo {author} {\bibfnamefont{S.~T.}\ \bibnamefont{Thompson}}
  \emph{et~al.},\ }%
  \bibfield{journal}{%
  \bibinfo {journal} {Phys. Rev. Lett.}\ }%
  \textbf{\bibinfo {volume} {95}},\ \bibinfo {pages} {190404} (\bibinfo {year}
  {2005})%
  \bibAnnoteFile{NoStop}{thompson:2005}%
\bibitem{zhang:2009}%
  \BibitemOpen
  \bibfield{author}{%
  \bibinfo {author} {\bibfnamefont{P.}~\bibnamefont{Zhang}} \emph{et~al.},\ }%
  \bibfield{journal}{%
  \bibinfo {journal} {Phys. Rev. Lett.}\ }%
  \textbf{\bibinfo {volume} {103}},\ \bibinfo {pages} {133202} (\bibinfo {year}
  {2009})%
  \bibAnnoteFile{NoStop}{zhang:2009}%
\bibitem{kaufman:2009}%
  \BibitemOpen
  \bibfield{author}{%
  \bibinfo {author} {\bibfnamefont{A.}~\bibnamefont{Kaufman}} \emph{et~al.},\
  }%
  \bibfield{journal}{%
  \bibinfo {journal} {Phys. Rev. A}\ }%
  \textbf{\bibinfo {volume} {80}},\ \bibinfo {pages} {050701} (\bibinfo {year}
  {2009})%
  \bibAnnoteFile{NoStop}{kaufman:2009}%
\bibitem{verhaar:2009}%
  \BibitemOpen
  \bibfield{author}{%
  \bibinfo {author} {\bibfnamefont{B.}~\bibnamefont{Verhaar}} \emph{et~al.},\
  }%
  \bibfield{journal}{%
  \bibinfo {journal} {Phys. Rev. A}\ }%
  \textbf{\bibinfo {volume} {79}},\ \bibinfo {pages} {032711} (\bibinfo {year}
  {2009})%
  \bibAnnoteFile{NoStop}{verhaar:2009}%
\bibitem{cohen:1992}%
  \BibitemOpen
  \bibfield{author}{%
  \bibinfo {author} {\bibfnamefont{C.}~\bibnamefont{Cohen-Tannoudji}}, \bibinfo
  {author} {\bibfnamefont{J.}~\bibnamefont{Dupont-Roc}},\ and\ \bibinfo
  {author} {\bibfnamefont{G.}~\bibnamefont{Grynberg}},\ }%
  \emph{\bibinfo {title} {Atom-Photon Interactions}}\ (\bibinfo {publisher}
  {Wiley},\ \bibinfo {address} {New York},\ \bibinfo {year} {1992})%
  \bibAnnoteFile{NoStop}{cohen:1992}%
\bibitem{Note1}%
  \BibitemOpen
  \bibinfo {note} {This restriction leads to simpler algebra, but is not
  essential: any polarization can be decomposed into $\sigma _\pm $ components
  and, for a given $\omega $, only one of the components will induce the
  desired resonant coupling to a bound state.}%
  \bibAnnoteFile{Stop}{Note1}%
\bibitem{Note2}%
  \BibitemOpen
  \bibinfo {note} {In Eq.~\ref {eq:couplingW} we omit a small coupling of the
  mw to the nuclear spins which does not affect the results.}%
  \bibAnnoteFile{Stop}{Note2}%
\bibitem{Note3}%
  \BibitemOpen
  \bibinfo {note} {For $s$--wave collisions between bosons, only symmetric
  internal states are relevant.}%
  \bibAnnoteFile{Stop}{Note3}%
\bibitem{stoof:1988}%
  \BibitemOpen
  \bibfield{author}{%
  \bibinfo {author} {\bibfnamefont{H.~T.}\ \bibnamefont{Stoof}} \emph{et~al.},\
  }%
  \bibfield{journal}{%
  \bibinfo {journal} {Phys. Rev. B}\ }%
  \textbf{\bibinfo {volume} {38}},\ \bibinfo {pages} {4688} (\bibinfo {year}
  {1988})%
  \bibAnnoteFile{NoStop}{stoof:1988}%
\bibitem{verhaar:1993}%
  \BibitemOpen
  \bibfield{author}{%
  \bibinfo {author} {\bibfnamefont{B.}~\bibnamefont{Verhaar}} \emph{et~al.},\
  }%
  \bibfield{journal}{%
  \bibinfo {journal} {Phys. Rev. A}\ }%
  \textbf{\bibinfo {volume} {48}},\ \bibinfo {pages} {R3429} (\bibinfo {year}
  {1993})%
  \bibAnnoteFile{NoStop}{verhaar:1993}%
\bibitem{nr3:2007}%
  \BibitemOpen
  \bibfield{author}{%
  \bibinfo {author} {\bibfnamefont{W.}~\bibnamefont{Press}} \emph{et~al.},\ }%
  \emph{\bibinfo {title} {Numerical Recipes}}\ (\bibinfo {publisher} {CUP},\
  \bibinfo {year} {2007})\ Chap.~\bibinfo {chapter} {18}%
  \bibAnnoteFile{NoStop}{nr3:2007}%
\bibitem{Note4}%
  \BibitemOpen
  \bibinfo {note} {The same approaches were used for method (ii).}%
  \bibAnnoteFile{Stop}{Note4}%
\bibitem{arimondo:1977}%
  \BibitemOpen
  \bibfield{author}{%
  \bibinfo {author} {\bibfnamefont{E.}~\bibnamefont{Arimondo}} \emph{et~al.},\
  }%
  \bibfield{journal}{%
  \bibinfo {journal} {Rev. Mod. Phys.}\ }%
  \textbf{\bibinfo {volume} {49}},\ \bibinfo {pages} {31} (\bibinfo {year}
  {1977})%
  \bibAnnoteFile{NoStop}{arimondo:1977}%
\bibitem{barakat:1986}%
  \BibitemOpen
  \bibfield{author}{%
  \bibinfo {author} {\bibfnamefont{B.}~\bibnamefont{Barakat}} \emph{et~al.},\
  }%
  \bibfield{journal}{%
  \bibinfo {journal} {Chemical Physics}\ }%
  \textbf{\bibinfo {volume} {102}},\ \bibinfo {pages} {215 } (\bibinfo {year}
  {1986})%
  \bibAnnoteFile{NoStop}{barakat:1986}%
\bibitem{yan:1996}%
  \BibitemOpen
  \bibfield{author}{%
  \bibinfo {author} {\bibfnamefont{Z.-C.}\ \bibnamefont{Yan}} \emph{et~al.},\
  }%
  \bibfield{journal}{%
  \bibinfo {journal} {Phys. Rev. A}\ }%
  \textbf{\bibinfo {volume} {54}},\ \bibinfo {pages} {2824} (\bibinfo {year}
  {1996})%
  \bibAnnoteFile{NoStop}{yan:1996}%
\bibitem{colavecchia:2003}%
  \BibitemOpen
  \bibfield{author}{%
  \bibinfo {author} {\bibfnamefont{F.}~\bibnamefont{Colavecchia}}
  \emph{et~al.},\ }%
  \bibfield{journal}{%
  \bibinfo {journal} {J. Chem. Phys.}\ }%
  \textbf{\bibinfo {volume} {118}},\ \bibinfo {pages} {5484} (\bibinfo {year}
  {2003})%
  \bibAnnoteFile{NoStop}{colavecchia:2003}%
\bibitem{vanabeelen:1999}%
  \BibitemOpen
  \bibfield{author}{%
  \bibinfo {author} {\bibfnamefont{F.~A.}\ \bibnamefont{van Abeelen}}\ and\
  \bibinfo {author} {\bibfnamefont{B.~J.}\ \bibnamefont{Verhaar}},\ }%
  \bibfield{journal}{%
  \bibinfo {journal} {Phys. Rev. A}\ }%
  \textbf{\bibinfo {volume} {59}},\ \bibinfo {pages} {578} (\bibinfo {year}
  {1999})%
  \bibAnnoteFile{NoStop}{vanabeelen:1999}%
\bibitem{falke:2008}%
  \BibitemOpen
  \bibfield{author}{%
  \bibinfo {author} {\bibfnamefont{S.}~\bibnamefont{Falke}} \emph{et~al.},\ }%
  \bibfield{journal}{%
  \Doi{10.1103/PhysRevA.78.012503}{\bibinfo {journal} {Phys. Rev. A}}\ }%
  \textbf{\bibinfo {volume} {78}},\ \bibinfo {pages} {012503} (\bibinfo {year}
  {2008})%
  \bibAnnoteFile{NoStop}{falke:2008}%
\bibitem{kokkelmans:2008}%
  \BibitemOpen
  \bibfield{author}{%
  \bibinfo {author} {\bibfnamefont{S.}~\bibnamefont{Kokkelmans}},\ }%
  \bibinfo {howpublished} {private communication} (\bibinfo {year} {2008})%
  \bibAnnoteFile{NoStop}{kokkelmans:2008}%
\bibitem{leo:2000}%
  \BibitemOpen
  \bibfield{author}{%
  \bibinfo {author} {\bibfnamefont{P.~J.}\ \bibnamefont{Leo}} \emph{et~al.},\
  }%
  \bibfield{journal}{%
  \Doi{10.1103/PhysRevLett.85.2721}{\bibinfo {journal} {Phys. Rev. Lett.}}\ }%
  \textbf{\bibinfo {volume} {85}},\ \bibinfo {pages} {2721} (\bibinfo {year}
  {2000})%
  \bibAnnoteFile{NoStop}{leo:2000}%
\bibitem{amiot:2002}%
  \BibitemOpen
  \bibfield{author}{%
  \bibinfo {author} {\bibfnamefont{C.}~\bibnamefont{Amiot}}\ and\ \bibinfo
  {author} {\bibfnamefont{O.}~\bibnamefont{Dulieu}},\ }%
  \bibfield{journal}{%
  \bibinfo {journal} {J. Chem. Phys.}\ }%
  \textbf{\bibinfo {volume} {117}},\ \bibinfo {pages} {5155} (\bibinfo {year}
  {2002})%
  \bibAnnoteFile{NoStop}{amiot:2002}%
\bibitem{chin:2004}%
  \BibitemOpen
  \bibfield{author}{%
  \bibinfo {author} {\bibfnamefont{C.}~\bibnamefont{Chin}} \emph{et~al.},\ }%
  \bibfield{journal}{%
  \Doi{10.1103/PhysRevA.70.032701}{\bibinfo {journal} {Phys. Rev. A}}\ }%
  \textbf{\bibinfo {volume} {70}},\ \bibinfo {pages} {032701} (\bibinfo {year}
  {2004})%
  \bibAnnoteFile{NoStop}{chin:2004}%
\bibitem{vanhaecke:2004}%
  \BibitemOpen
  \bibfield{author}{%
  \bibinfo {author} {\bibfnamefont{N.}~\bibnamefont{Vanhaecke}} \emph{et~al.},\
  }%
  \bibfield{journal}{%
  \bibinfo {journal} {Eur. Phys. J. D}\ }%
  \textbf{\bibinfo {volume} {28}},\ \bibinfo {pages} {351} (\bibinfo {year}
  {2004})%
  \bibAnnoteFile{NoStop}{vanhaecke:2004}%
\bibitem{xie:2009}%
  \BibitemOpen
  \bibfield{author}{%
  \bibinfo {author} {\bibfnamefont{F.}~\bibnamefont{Xie}} \emph{et~al.},\ }%
  \bibfield{journal}{%
  \bibinfo {journal} {J. Chem. Phys.}\ }%
  \textbf{\bibinfo {volume} {130}},\ \bibinfo {pages} {051102} (\bibinfo {year}
  {2009})%
  \bibAnnoteFile{NoStop}{xie:2009}%
\bibitem{abraham:1997}%
  \BibitemOpen
  \bibfield{author}{%
  \bibinfo {author} {\bibfnamefont{E.~R.~I.}\ \bibnamefont{Abraham}}
  \emph{et~al.},\ }%
  \bibfield{journal}{%
  \bibinfo {journal} {Phys. Rev. A}\ }%
  \textbf{\bibinfo {volume} {55}},\ \bibinfo {pages} {R3299} (\bibinfo {year}
  {1997})%
  \bibAnnoteFile{NoStop}{abraham:1997}%
\bibitem{soding:1998}%
  \BibitemOpen
  \bibfield{author}{%
  \bibinfo {author} {\bibfnamefont{J.}~\bibnamefont{S{\"o}ding}}
  \emph{et~al.},\ }%
  \bibfield{journal}{%
  \bibinfo {journal} {Phys. Rev. Lett.}\ }%
  \textbf{\bibinfo {volume} {80}},\ \bibinfo {pages} {1869} (\bibinfo {year}
  {1998})%
  \bibAnnoteFile{NoStop}{soding:1998}%
\bibitem{papoular:tbp}%
  \BibitemOpen
  \bibfield{author}{%
  \bibinfo {author} {\bibfnamefont{D.~J.}\ \bibnamefont{Papoular}}
  \emph{et~al.},\ }%
  \bibinfo {note} {to be published}%
  \bibAnnoteFile{NoStop}{papoular:tbp}%
\end{thebibliography}
%Merlin.mbs v4.21 2009-07-09.
%

\end{document}